\documentclass[superscriptaddress,aps,preprintnumbers,amsmath,amssymb,prd,nofootinbib,preprint]{revtex4-1}
\usepackage{setspace}
\usepackage{graphicx}
\usepackage{amsfonts}
\usepackage[mathscr]{eucal}
\usepackage{ascmac}
\usepackage{epstopdf}
\usepackage{dcolumn}% Align table columns on decimal point
\usepackage{bm}% bold math
\usepackage{hyperref}
\usepackage{color}
\usepackage{cancel}
\begin{document}

%%%%%%%%%%%%%%%%%%%%%%%%%%%%%%%%%%%%%%%%%%%

\def\a{\alpha}
\def\b{\beta}
\def\c{\varepsilon}
\def\d{\delta}
\def\e{\epsilon}
\def\f{\phi}
\def\g{\gamma}
\def\h{\theta}
\def\k{\kappa}
\def\l{\lambda}
\def\m{\mu}
\def\n{\nu}
\def\p{\psi}
\def\q{\partial}
\def\r{\rho}
\def\s{\sigma}
\def\t{\tau}
\def\u{\upsilon}
\def\v{\varphi}
\def\w{\omega}
\def\x{\xi}
\def\y{\eta}
\def\z{\zeta}
\def\D{\Delta}
\def\G{\Gamma}
\def\H{\Theta}
\def\L{\Lambda}
\def\F{\Phi}
\def\P{\Psi}
\def\S{\Sigma}

\def\o{\over}
\def\beq{\begin{eqnarray}}
\def\eeq{\end{eqnarray}}
\newcommand{\gsim}{ \mathop{}_{\textstyle \sim}^{\textstyle >} }
\newcommand{\lsim}{ \mathop{}_{\textstyle \sim}^{\textstyle <} }
\newcommand{\vev}[1]{ \left\langle {#1} \right\rangle }
\newcommand{\bra}[1]{ \langle {#1} | }
\newcommand{\ket}[1]{ | {#1} \rangle }
\newcommand{\EV}{ {\rm eV} }
\newcommand{\KEV}{ {\rm keV} }
\newcommand{\MEV}{ {\rm MeV} }
\newcommand{\GEV}{ {\rm GeV} }
\newcommand{\TEV}{ {\rm TeV} }
\newcommand{\1}{\mbox{1}\hspace{-0.25em}\mbox{l}}
\def\diag{\mathop{\rm diag}\nolimits}
\def\Spin{\mathop{\rm Spin}}
\def\SO{\mathop{\rm SO}}
\def\O{\mathop{\rm O}}
\def\SU{\mathop{\rm SU}}
\def\U{\mathop{\rm U}}
\def\Sp{\mathop{\rm Sp}}
\def\SL{\mathop{\rm SL}}
\def\tr{\mathop{\rm tr}}

\def\IJMP{Int.~J.~Mod.~Phys. }
\def\MPL{Mod.~Phys.~Lett. }
\def\NP{Nucl.~Phys. }
\def\PL{Phys.~Lett. }
\def\PR{Phys.~Rev. }
\def\PRL{Phys.~Rev.~Lett. }
\def\PTP{Prog.~Theor.~Phys. }
\def\ZP{Z.~Phys. }

\def\dd{\mathrm{d}}
\def\ff{\mathrm{f}}
\def\BH{{\rm BH}}
\def\inf{{\rm inf}}
\def\ev{{\rm evap}}
\def\eq{{\rm eq}}
\def\SM{{\rm sm}}
\def\Mpl{M_{\rm Pl}}
\def\GeV{{\rm GeV}}
\newcommand{\Red}[1]{\textcolor{red}{#1}}

\def\mDM{m_{\rm DM}}
\def\mphi{m_{\phi}}
\def\TeV{{\rm TeV}}
\def\Gphi{\Gamma_\phi}
\def\TR{T_{\rm RH}}
\def\Br{{\rm Br}}
\def\DM{{\rm DM}}
\def\Eth{E_{\rm th}}
\newcommand{\lmk}{\left(}  
\newcommand{\rmk}{\right)}
\newcommand{\lkk}{\left[}  
\newcommand{\rkk}{\right]}
\newcommand{\lhk}{\left \{ }  
\newcommand{\rhk}{\right \} }
\newcommand{\del}{\partial}  
\newcommand{\la}{\left\langle} 
\newcommand{\ra}{\right\rangle}

%% DIRAC SLASH %%%%%%%%%%%%%%%%%%%
\def\slashchar#1{\setbox0=\hbox{$#1$} % set a box for #1
\dimen0=\wd0 % and get its size
\setbox1=\hbox{/} \dimen1=\wd1 % get size of /
\ifdim\dimen0>\dimen1 % #1 is bigger
\rlap{\hbox to \dimen0{\hfil/\hfil}} % so center / in box
#1 % and print #1
\else % / is bigger
\rlap{\hbox to \dimen1{\hfil$#1$\hfil}} % so center #1
/ % and print /
\fi}

%%%%%%%%%%%%%%%%%%%%%%%%%%%%%%%%%%%%%%%%%%%%%%%%%%%%%%%%%%%%%%%

\title{
Anomaly Mediated Gaugino Mass and Path-Integral Measure
}

\author{Keisuke Harigaya}
\affiliation{Kavli IPMU (WPI), TODIAS, University of Tokyo, Kashiwa, 277-8583, Japan}
\author{Masahiro Ibe}
\affiliation{ICRR, University of Tokyo, Kashiwa, 277-8582, Japan}
\affiliation{Kavli IPMU (WPI), TODIAS, University of Tokyo, Kashiwa, 277-8583, Japan}
\begin{abstract}
In recent years, there have been controversy concerning the anomaly mediated gaugino
mass in the superspace formalism of supergravity.
In this paper, we reexamine the gaugino mass term in this formalism by paying particular 
attention to symmetry which controls gaugino masses in supergravity.
We first discuss super-Diffeomorphism invariance of path-integral measures of charged superfields.
As we will show, the super-Diffeomorphism invariant measure is not invariant under 
a super-Weyl transformation,
which is the origin of the anomaly mediated gaugino mass.
We show how the anomaly mediated gaugino mass is expressed as a local operator 
in a Wilsonian effective action in a super-Diffeomorphism covariant way.
We also obtain a gaugino mass term independent of the gauge choice of the fictitious 
super-Weyl symmetry in the super-Weyl compensator formalism,
which reproduces the widely accepted result.
Besides, we discuss how to reconcile the gaugino mass term in the local Wilsonian effective action and 
the gaugino mass term appearing in a non-local 1PI quantum effective action.
\end{abstract}

\date{\today}
\maketitle
\preprint{IPMU 14-0303}
\preprint{ICRR-report-693-2014-19}
%%%%%%%%%%%%%%%%%%%%%%%%%%%%%%%%%%%%%%%%%%%%%%%%%%%%%%%%%%%%

%%%%%%%%%%%%%%%%%%%%%%%%%%%%%%%%%%%%%%%%%%%%%%%%%%%%%%%%%%%%
\section{Introduction}
After the discovery of the Higgs boson with a mass about $126$\,GeV at the LHC experiments\,\cite{Aad:2012tfa},
the anomaly mediation mechanism for the gaugino mass generation\,\cite{Giudice:1998xp,Randall:1998uk} 
(see also \cite{Dine:1992yw}) is gathering renewed attention.
The anomaly mediated gaugino mass plays a crucial role in constructing a class 
of high scale supersymmetry models where sfermion masses are 
in a hundreds to thousands TeV range while gaugino masses are within a TeV range 
as proposed in \cite{Giudice:1998xp,Wells:2004di} and subsequently in 
\cite{Ibe:2006de,Acharya:2007rc,Hall:2011jd,Ibe:2011aa,Bhattacherjee:2012ed,
Dudas:2012wi,Arvanitaki:2012ps,ArkaniHamed:2012gw,Evans:2013lpa,Harigaya:2013asa,Evans:2013nka}. 
On top of a successful prediction on the observed Higgs boson mass\,\cite{TU-363,Giudice:2004tc,ArkaniHamed:2004yi},
models in this class are free from the so-called cosmological Polonyi problem\,\cite{Coughlan:1983ci}
(see also \cite{Ibe:2006am}),  for no singlet supersymmetry breaking fields are required in the models.%
\footnote{A simple implementation of the $\mu$-term in this class of ``without singlet" models
was first done in \cite{Ibe:2006de} by coupling the Higgs doublets to the $R$-symmetry breaking sector
along the lines of the Casas-Munoz mechanism\,\cite{Casas:1992mk} 
(or the generalization of the Giudice-Masiero mechanism\,\cite{Giudice:1988yz}).
}
This class of models is also free from infamous gravitino problems\,\cite{Pagels:1981ke,kkm}.
Besides, the lightest supersymmetric particle is the almost pure Wino in most parameter space, 
which is a good candidate for dark matter when it is produced by either 
thermally\,\cite{Hisano:2006nn,Cirelli:2007xd} 
or non-thermally\,\cite{Moroi:1999zb,Gherghetta:1999sw} 
(see also \cite{Ibe:2004tg}).%
\footnote{
For the current status and future prospects of Wino dark matter detection, see \cite{Cohen:2013ama,Fan:2013faa,Hryczuk:2014hpa,Bhattacherjee:2014dya}, and for collider searches see \cite{Cirelli:2014dsa}.  
See also \cite{Ibe:2012hr,Ibe:2012sx,Harigaya:2014dwa,Harigaya:2014pqa} 
for related discussions on wino dark matter.}

As illustrated in \cite{Giudice:1998xp,Randall:1998uk}, 
the most transparent way to look at the anomaly mediated gaugino mass 
is to use the conformal compensator formalism of supergravity\,\cite{Siegel:1978mj,Kugo:1982cu}.
In this formalism, only the conformal compensator has a non-vanishing 
 $F$-term vacuum expectation value (VEV) after supersymmetry breaking in the gravity sector, 
and hence, the anomaly mediated gaugino mass can be 
extracted by looking at how the chiral compensator appears in the gauge kinetic function.
(See also \cite{Bagger:1999rd,Boyda:2001nh,Dine:2007me,Jung:2009dg,Sanford:2010hc,DEramo:2012qd,D'Eramo:2013mya,Dine:2013nka,DiPietro:2014moa} 
for informative discussions on the anomaly mediated gaugino mass.)

In the last few years, however, there has been controversy 
\cite{deAlwis:2008aq,D'Eramo:2013mya} over how the anomaly-mediated gaugino mass
appears in the superspace formalism of supergravity\,\cite{Wess:1992cp,Gates:1983nr}.
In particular, the author of \cite{deAlwis:2008aq} examined the gaugino mass in this formalism 
by introducing a chiral super-Weyl compensator field, $C$, in the track of \cite{Kaplunovsky:1994fg}, 
so that the model possesses a fictitious (but exact) super-Weyl gauge symmetry. 
Then, by looking at how the super-Weyl compensator $C$ appears in the gauge kinetic function,
the author claimed that the anomaly mediated gaugino mass derived in \,\cite{Giudice:1998xp,Randall:1998uk}
vanishes.
This claim was refuted by a subsequent paper\,\cite{D'Eramo:2013mya}, which pointed out that 
the gravity multiplets also possess non-vanishing $F$-term VEVs for the gauge choice of 
the fictitious super-Weyl gauge symmetry in \cite{deAlwis:2008aq}.
Thus, the full anomaly-mediated gaugino masses cannot be extracted just by looking
at the $C$ dependence of the gauge kinetic function.
Eventually, by arguing that the gaugino mass should be independent of 
the gauge choice of the super-Weyl gauge symmetry, the gaugino mass in \,\cite{Giudice:1998xp,Randall:1998uk} 
is reproduced in \cite{D'Eramo:2013mya} by taking a gauge in which the $F$-term VEVs
of the gravity multiplets vanish.

In these discussions,
there remain unsettled questions.
First of all, it is not clear whether the anomaly mediated gaugino mass can be expressed
as a local operator in a Wilsonian effective action in a super-Diffeomorphism covariant way 
without invoking the super-Weyl symmetry compensator $C$.
Second, the lack of the local term expression without $C$ inevitably seems to mean 
that there is no consistent expression of the anomaly mediated gaugino mass as a 
local operator independent of the gauge choice of the fictitious super-Weyl gauge symmetry.

The main purpose of this paper is to settle these problems.
For that purpose, we reexamine the gaugino mass term in the superspace formalism of supergravity 
without invoking the fictitious super-Weyl gauge symmetry.
We pay particular attention to super-Diffeomorphism invariance of path-integral measures of
charged supermultiplets.
As we will show, the super-Diffeomorphism invariant measure is not invariant under 
an approximate super-Weyl symmetry which
forbids the gaugino mass
at the classical level.
Anomalous breaking of the approximate super-Weyl symmetry 
(not to be confused with the fictitious super-Weyl symmetry in \cite{Kaplunovsky:1994fg})
is the origin of the anomaly mediated gaugino mass in the superspace formalism.

Armed with the super-Diffeomorphism invariant measure, we show that the anomaly mediated gaugino mass 
can be read off from a local operator in a Wilsonian action when we change the path-integral measure
from the super-Diffeomorphism invariant one to the super-Weyl invariant one.
There, we emphasize that the corresponding local operator is not invariant under the super-Diffeomrophism.
The non-invariance of the local term is required for the super-Diffeomorphism invariance of the quantum theory.

Once we learn how the local gaugino mass term arises in the superspace formalism of supergravity,
it is straightforward to derive the local term expression of the gaugino mass term 
which is independent of the gauge choice of the fictitious super-Weyl gauge symmetry in the super-Weyl compensator formalism.
We also discuss how to reconcile 
the anomaly mediated gaugino mass term in the Wilsonian effective action
and the non-local expression of the gaugino mass term appearing in the 1PI quantum effective action
derived in \cite{Bagger:1999rd}.

The organization of this paper is as follows.
In Sec.\,\ref{sec:classical}, we discuss the gaugino mass appearing in the supergravity action
at the classical level. 
There, we show that the gaugino mass is highly suppressed at the classical level 
due to the approximate super-Weyl symmetry which is
respected by relevant interactions of gauge and charged matter supermultiplets.%
\footnote{Throughout this paper, relevant interaction terms denote the interaction terms with mass dimensions
less than or equal to $4$.}
In Sec.\,\ref{sec:measure}, we discuss the super-Diffeomorphism invariant path-integral measure of
the charged matter which has a non-trivial but unique dependence on the chiral density of the gravity multiplet.
There, we see that the super-Diffeomorphism invariant measure is not invariant under the 
approximate super-Weyl symmetry.
This property is important to understand how the anomaly mediated gaugino mass term can be 
expressed as a local term in the Wilsonian effective action in a super-Diffeomorphism covariant way.
In Sec.\,\ref{sec:fictitious SW}, we show the local expression of the gaugino mass term
which is independent of the gauge choice of the fictitious super-Weyl gauge symmetry in the super-Weyl compensator formalism 
in \cite{Kaplunovsky:1994fg}.
We also show how the gaugino mass term is related to the gaugino mass term 
in the non-local 1PI quantum effective action derived in \cite{Bagger:1999rd}.
We summarize our discussion in Sec.\,\ref{sec:summary}.

\label{sec:intro}
%%%%%%%%%%%%%%%%%%%%%%%%%%%%%%%%%%%%%%%%%%%%%%%%%%%%%%%%%%%%%%%%

%%%%%%%%%%%%%%%%%%%%%%%%%%%%%%%%%%%%%%%%%%%%%%%%%%%%%%%%%%%%%%%%
\section{Approximate super-Weyl symmetry in classical action}
\label{sec:classical}
%%%%%%%%%%%%%%%%%%%%%%%%%%%%%%%%%%%%%%%%%%%%%%%%%%%%%%%%%%%%%%%%
Before discussing the anomaly mediated gaugino mass, let us first clarify the
gaugino mass expected in the local supergravity action at the classical level.%
\footnote{Here, we assume that the classical action consists of local interactions.
If the classical action is allowed to be non-local, an arbitrary gaugino mass of $O(m_{3/2})$ 
can be introduced by using the non-local term in Eq.~(\ref{eq:non-local}) without 
conflicting with the super-Diffeomorphism invariance.
}
In our discussion, we concentrate ourselves in a situation where 
supersymmetry is dominantly broken by some charged fields under some symmetries or by composite fields.
Otherwise direct interactions between the supersymmetry breaking fields and gauge multiplets 
lead to the ``tree-level" gaugino mass of the order of the gravitino mass, $m_{3/2}$.
Under this assumption, the direct interactions between the supersymmetry 
breaking fields and the gauge supermultiplets are suppressed at least by a second power of 
the Planck scale, $M_{\rm PL}$, and hence, resultant gaugino masses 
from those interactions are negligible.
By the same reason, we also assume that no supersymmetry breaking field obtains a vacuum 
expectation value of the order of the Planck scale.%
\footnote{These assumptions also reduce contributions to 
gaugino masses from the K\"ahler and sigma-model anomalies\,\cite{Bagger:1999rd,DEramo:2012qd}.
}

Once we assume that the gaugino mass from couplings to the supersymmetry breaking 
sector is highly suppressed, remaining sources of the gaugino mass are couplings
to the supergravity multiplets.
As is well known, however, gaugino masses from tree level interactions 
to the supergravity multiplets are also suppressed in spite of the apparent $F$-term VEVs of $O(m_{3/2})$ 
in the supergravity multiplets. 
As we shortly discuss, the absence of $O(m_{3/2})$ gaugino masses from the supergravity multiplets is due to an approximate super-Weyl symmetry,
which is the key to understand the origin of the anomaly mediated gaugino mass in the next section.
For the time being, we restrict ourselves to the gaugino mass generation in a $U(1)$ gauge theory 
with a pair of vector-like matters.
The following discussion can be extended to general non-abelian gauge theories (see discussions
in Sec.\,\ref{sec:summary}).

\subsection{Classical supergravity action}
In this paper, we follow the notation and the formulation in \cite{Wess:1992cp}, 
except for the notation of complex conjugate (we use $\dagger$) and for
the normalization of gauge supermultiplets to which we adopt the one in \cite{ArkaniHamed:1997mj}.
For a simple model with charged chiral multiplets $Q$ and $\bar{Q}$, and an $U(1)$ gauge multiplet $V$, 
the classical supergravity action is given by,
\begin{eqnarray}
\label{eq:classical action}
{\cal L} =  \Mpl^2 \int {\rm d}^2 \Theta\, 2 {\cal E} \, \frac{3}{8} \left({\cal D}^{\dag2} - 8R\right){\rm exp}\left[-\frac{K}
{3\Mpl^2}\right]
+ 
\frac{1}{16g^2}\int {\rm d}^2 \Theta\, 2{\cal E} \,W^\alpha W_\alpha + {\rm h.c.},\nonumber \\
K
= Q^\dag e^{2V} Q + \bar{Q}^\dag e^{-2V} \bar{Q} +\cdots,~~
W_\alpha \equiv -\frac{1}{4} \left({\cal D}^{\dag 2}- 8R\right) \left(
e^{-2V} {\cal D}_\alpha e^{2V}
\right),
\end{eqnarray}
where $\Theta^\alpha$, ${\cal E}$, ${\cal D}_{\alpha}$, $R$, $K$, and $g$ are the fermionic coordinate, the chiral density, the covariant derivative, the superspace curvature, the K\"ahler potential, and the gauge coupling constant, respectively.
Here, we have assumed that the chiral multiplets $Q$ and $\bar{Q}$ are massless.
By expanding the chiral multiplets, we can extract relevant  interactions,
\begin{eqnarray}
\label{eq:matter kin}
{\cal L}_{\rm kin,matter} &=&  -\frac{1}{8} \int {\rm d}^2 \Theta\,2 {\cal E} \left({\cal D}^{\dag2} - 8R\right) \left( Q^\dag e^{2V} Q + \bar{Q}^\dag e^{-2V} \bar{Q} \right) + {\rm h.c} \ ,\\
\label{eq:gauge kin}
{\cal L}_{\rm kin,gauge} &=& \frac{1}{16g^2}\int {\rm d}^2 \Theta \,2{\cal E}\, W^\alpha W_\alpha + {\rm h.c.}\ ,
\end{eqnarray}
from which we can extract gauge interactions and kinetic terms.
Other interactions are suppressed by the Planck scale.

Now, let us expand $W^\alpha$, ${\cal E}$, and $R$ in terms of component fields;
\begin{eqnarray}
\label{eq:expansion}
W^\alpha &=& - 2 i \lambda^\alpha + \cdots \nonumber\ , \\
2 {\cal E} & = & e( 1-M^* \Theta^2) + \cdots\nonumber\ ,\\
R & = & -\frac{1}{6} M - \frac{1}{9} |M|^2\Theta^2 + \cdots .
\end{eqnarray}
Here, $\lambda^\alpha$, $e$, and $M$ are the gaugino, the determinant of the vielbein, 
and the auxiliary scalar component of the gravity multiplet, respectively. 
The ellipses denote terms which are irrelevant for our discussion on the gaugino mass.
The auxiliary field $M$ is fixed by the equation of motion as
\begin{eqnarray}
 M^*  = - 3 m_{3/2}\ ,
 \label{eq:auxM}
\end{eqnarray}
where we have omitted contributions from the supersymmetry breaking sector which
are negligible under the assumption we have made at the beginning of this section.

Since the chiral density ${\cal E}$ has a non-vanishing $\Theta^2$ term, 
it might look non-trivial why the gaugino mass of $O(m_{3/2})$ does not appear 
from the interaction in Eq.~(\ref{eq:gauge kin}).
In the rest of this section, we show that the absence of the gaugino mass in the classical action is understood by an approximate super-Weyl symmetry.

\subsection{Approximate super-Weyl symmetry}
Let us consider the super-Weyl transformation parameterized by a chiral scalar $\Sigma$~\cite{Wess:1992cp},%
\footnote{In this paper we define an infinitesimal transformation of a superfield $\cal X$ by
${\cal X}' ={\cal X} -\delta {\cal X} $.}
\begin{eqnarray}
\label{eq:SW}
\delta_{\rm SW} {\cal E} &=& 6 \Sigma {\cal E} + \frac{\partial}{\partial \Theta^\alpha}\left(S^\alpha {\cal E}\right)\ , \nonumber\\
\delta_{\rm SW} R &=&  -4 \Sigma R  - \frac{1}{4}\left( {\cal D}^{\dag 2} - 8 R\right) \Sigma^\dag - S^\alpha \frac{\partial}{\partial \Theta^\alpha}R\ , \nonumber\\
\delta_{\rm SW} W^\alpha &=& - 3 \Sigma W^\alpha + \cdots\ ,\nonumber\\
\delta_{\rm SW} Q &=& w\Sigma Q - S^\alpha \frac{\partial}{\partial \Theta^\alpha}Q\ , \nonumber\\
S^\alpha &\equiv& \Theta^\alpha \left( 2 \Sigma^\dag - \Sigma \right)| +\Theta^2 {\cal D}^\alpha \Sigma|\ ,
\end{eqnarray}
where the ellipses denote terms which are irrelevant for our discussion. 
A parameter $w$ is the Weyl weight of $Q$.%
\footnote{If ${Q}$ is not a chiral scalar but a chiral density with a density weight $\tilde w$, 
the super-Weyl transformation is given by,
\begin{eqnarray}
\delta_{\rm SW} {Q} &=& w\Sigma {Q} - S^\alpha \frac{\partial}{\partial \Theta^\alpha}{Q}
+ \tilde{w} {Q}\frac{\partial}{\partial \Theta^\alpha}S^\alpha \ .
\end{eqnarray}
}
The $S^\alpha$-dependent terms are inhomogeneous transformations 
which can be cancelled by the super-Diffeomorphism (see Eq.\,(\ref{eq:Sdiff})).
From Eqs.\,(\ref{eq:expansion}) and (\ref{eq:SW}),  the transformation laws of $e$, $M$ and $\lambda^\alpha$ are 
given by
\begin{eqnarray}
\label{eq:Mlambda}
\delta_{\rm SW} e &=& 4 \left(\Sigma + \Sigma^\dag\right)| e, \nonumber \\
\delta_{\rm SW} M &=& -2 (2 \Sigma - \Sigma^\dag)|M + \frac{3}{2} {\cal D}^{\dag 2}\Sigma^\dag|, \nonumber\\
\delta_{\rm SW} \lambda^\alpha &=& - 3 \Sigma| \lambda^\alpha,
\end{eqnarray}
where ${\cal X}|$ denotes the lowest component of a superfield $\cal{X}$.

From the transformation laws of the component fields in Eq.\,(\ref{eq:Mlambda}), it is clear that the possible
origin of the gaugino mass of $O(m_{3/2})$, 
\begin{eqnarray}
\int {\rm d}^4 x\, e M^{(*)} \lambda \lambda, 
\label{eq:MLL}
\end{eqnarray}
is not invariant under the super-Weyl transformation.
This shows that the gaugino mass is generated only through terms which break the super-Weyl symmetry.

As we immediately see, the kinetic term of the gauge multiplet in Eq.\,(\ref{eq:gauge kin}) 
is invariant under the super-Weyl transformation, and hence, does not contribute to the gaugino mass.
Higher dimensional terms omitted in Eq.\,(\ref{eq:classical action}) are, on the other hand,
not invariant under the super-Weyl transformation.
Contributions from such terms to the gaugino mass are, however, at the largest of $O(m_{3/2}^2/\Mpl)$, 
and hence are negligible.
Altogether, we find that there is no gaugino mass of $O(m_{3/2})$ from couplings to the supergravity multiplets
due to the approximate super-Weyl symmetry.%
\footnote{The term in Eq.\,(\ref{eq:MLL}) is invariant under the $R$-symmetry and the dilatational symmetry, parts of the super-Weyl symmetry.
Thus, the gaugino mass from the couplings to the supergravity multiplets 
cannot be forbidden by the $R$-symmetry nor the dilataional symmetry.}

For later convenience, let us also note that the terms of massless matter fields in Eq.~(\ref{eq:matter kin})
are also invariant under the super-Weyl symmetry.
That is, for $w = -2$, it can be shown that
\begin{eqnarray}
\label{eq:SW kin}
\delta_{\rm SW} \left( \left(\bar{{\cal D}}^2 - 8R\right) \left(Q^\dag  Q\right) \right) = -6 \Sigma  \left(\bar{{\cal D}}^2 - 8R\right) \left( Q^\dag  Q \right) -S^\alpha \frac{\partial}{\partial \Theta^\alpha} \left( \left(\bar{{\cal D}}^2 - 8R\right) \left(Q^\dag  Q\right) \right) \ .
\end{eqnarray}
From Eqs. (\ref{eq:SW}) and (\ref{eq:SW kin}),
the terms in Eq.~(\ref{eq:matter kin}) is invariant under the super-Weyl transformation.

Finally, let us stress that interaction terms of the gauge supermultiplets which are unsuppressed 
by the Planck scale is uniquely determined to the form of Eqs.~(\ref{eq:matter kin}) and~(\ref{eq:gauge kin}) by 
the super-Diffeomorphism invariance and by the  gauge invariance. 
Thus, one may regard the approximate super-Weyl symmetry as an accidental one.
Due to this accidental symmetry, the gaugino mass of $O(m_{3/2})$ is suppressed at the classical level.

%%%%%%%%%%%%%%%%%%%%%%%%%%%%%%%%%%%%%%%%%%%%%%%%%%%%%%%%%%%%%%%%
\section{Anomaly of the super-Weyl symmetry and Gaugino Mass}
\label{sec:measure}
%%%%%%%%%%%%%%%%%%%%%%%%%%%%%%%%%%%%%%%%%%%%%%%%%%%%%%%%%%
In the last section, we have shown that no gaugino mass of $O(m_{3/2})$ is generated through
couplings to the supergravity multiplets even after supersymmetry breaking
due to the approximate super-Weyl symmetry.
However, the approximate super-Weyl symmetry is in general broken by quantum effects.
In this section, we investigate effects of quantum violation of the approximate super-Weyl symmetry 
by Fujikawa's method\,\cite{Fujikawa:1979ay} in a Wilsonian effective action.

\subsection{Wilsonian effective action}
To discuss quantum effects on the super-Weyl symmetry, 
we take the local classical action in the previous section 
(Eq.\,(\ref{eq:classical action})) as the Wilsonian effective action with a cutoff at the Planck scale. 
Here, let us remind ourselves that effective quantum field theories suffer 
from ultraviolet divergences, and hence, they are well-defined only after the divergences are properly regularized.
In our arguments, we presume an ultraviolet regularization such that the ``tree-level" action
at the cutoff scale is manifestly invariant under the super-Diffeomorphism and the gauge transformations.
We refer to this super-Diffeomorphism invariant tree-level action at the cutoff scale as the Wilsonian effective action.%
\footnote{Although we fix the cutoff scale to the Planck scale for a while,
the following discussion is essentially unchanged as long as the cutoff scale 
is far larger than the gravitino mass. 
We also discuss effects of the change of the cutoff scale later.}

The Wilsonian effective action, in general, includes higher dimensional interactions than those 
in Eq.\,(\ref{eq:classical action}) suppressed by the cut off scale.
As we have discussed, however, contributions from those terms to the gaugino mass are highly suppressed
by the cutoff scale and hence negligible. 
One concern is whether non-local interaction terms appear in the Wilsonian effective action at the cutoff scale, 
which could lead to the gaugino mass of $O(m_{3/2})$.
In our argument, we presume that such non-local interactions do not show up in the Wilsonian effective action,
which is reasonable since we are dealing with effective field theories after integrating out ultraviolet modes.

\subsection{Super-Diffeomorphism invariance}
In the above definition of the super-Diffeomorphism invariant theory,
there is a missing ingredient, the measure of the path-integral. 
As elucidated in \cite{Fujikawa:1979ay}, the path-integral measure plays crucial role in discussing quantum
violations of symmetries.
Moreover, the definition of the ``tree-level" interactions in the Wilsonian effective action depends on the choice 
of the path-integral measure, which we will encounter shortly.
To clarify these issues, let us first discuss which path-integral measure we should use in conjunction with the ``tree-level" 
Wilsonian action.

Under the infinitesimal (chiral) super-Diffeomorphism transformation, $Q$
and ${\cal E}$ transform as 
\begin{eqnarray}
\label{eq:Sdiff}
Q \rightarrow Q'  &=& Q - \eta^M(x,\Theta)\partial _M Q\ ,\nonumber \\
{\cal E} \rightarrow {\cal E}' &=& {\cal E}- \eta^M(x,\Theta)\partial _M {\cal E} - (-)^M \left(\partial_M \eta^M\left(x, \Theta\right)\right){\cal E}\ ,
\end{eqnarray}
where $M = (m,\alpha)$ denotes the indices of the chiral super coordinate $(x^m,\Theta^\alpha)$,
$\eta^M (x, \Theta)$ parameterizes the super-Diffeomorphism, and
$(-)^M = (1, -1)$ for $M = (m,\alpha)$. 
As is shown in the Appendix~\ref{sec:Sdiff measure}, path-integral measures of chiral fields
are not invariant under the super-Diffeomorphism due to the anomaly of the gauge interactions,
i.e.
\begin{eqnarray}
 [D Q]  \to [D Q'] \neq  [D Q] \ , \quad
  [D \bar {Q}]  \to [D \bar{Q}'] \neq  [D \bar{Q}] \ .
 \label{eq:chiral measure}
\end{eqnarray}
Instead, anomaly free measures are given by
\begin{eqnarray}
\label{eq:SD measure}
[D\left(2{\cal E}\right)^{1/2}Q]
\ , \quad [D\left(2{\cal E}\right)^{1/2} \bar {Q}]\ .
\end{eqnarray}
For a later purpose, we define weighted chiral fields $Q_{\rm diff} = \left(2{\cal E}\right)^{1/2}Q$ ($\bar{Q}_{\rm diff} = \left(2{\cal E}\right)^{1/2}\bar{Q}$) which are no more a chiral scalar fields but chiral density fields 
with  density weights $1/2$.

In our discussion, we take the super-Diffeomorphism invariant Wilsonian 
effective action. 
Therefore, in order to obtain a super-Diffeomporphism invariant quantum theory,
we inevitably use the super-Diffeomorphism invariant path-integral measure in Eq.\,(\ref{eq:SD measure}).
If we use different measures, instead, we need to add appropriate super-Diffeomorphism variant counter terms 
to the tree-level Wilsonian action so that the super-Diffeomorphism is restored in the quantum theory.

\subsection{Anomalous breaking of the super-Weyl symmetry}
Once we choose appropriate path-integral measures for the charged fields, 
we can now discuss quantum violation of the super-Weyl symmetry.
Here, since we are interested in the gaugino mass, we only look at the breaking of the super-Weyl
symmetry by the anomaly of the corresponding gauge interaction.

Before proceeding further, let us comment on a technical point.
As   in Eq.\,(\ref{eq:SW}), the super-Weyl transformation 
is accompanied by a super-Diffeomorphism parameterised by $S^\alpha$, so that 
the super-Weyl symmetry can be expressed in terms of the component fields
defined in the chiral superspace spanned by $(x,\Theta)$.
The accompanied super-Diffeomorphism, however, makes it complicated to
discuss the quantum violation of the super-Weyl symmetry.
To avoid such a complication, we only consider a subset of the super-Weyl transformation
where $\Sigma$ has only an $F$-term, i.e.
\begin{eqnarray}
\Sigma(x,\Theta) = f(x)\Theta^2\ .
\end{eqnarray}
Here, $f$ is an arbitrary function of the space-time.
Under this restricted super-Weyl transformation, we find $S^\alpha = 0$, and hence,
no super-Diffeomorphism is accompanied.
We refer this type of the super-Weyl transformation  as an ``$F$-type" super-Weyl transformation.
It should be noted that the $F$-type super-Weyl transformation is
sufficient to forbid the gaugino mass from the term in Eq.\,(\ref{eq:gauge kin}) in
the discussion in Sec.\,\ref{sec:classical}.
In the followings, we concentrate on the anomalous breaking of the $F$-type super-Weyl symmetry.

Now let us examine the invariance of the path-integral measures in Eq.\,(\ref{eq:SD measure})
under the $F$-type super-Weyl transformation.
Under the transformation, $Q_{\rm diff}$ and $\bar{Q}_{\rm diff}$ are not invariant but transform by
\begin{eqnarray}
Q_{\rm diff} = \left(2{\cal E}\right)^{1/2}Q \to Q_{\rm diff}' = e^{-\Sigma} Q_{\rm diff}\ , 
\quad
\bar{Q}_{\rm diff} = \left(2{\cal E}\right)^{1/2}\bar{Q} \to \bar{Q}_{\rm diff}' = e^{-\Sigma} \bar{Q}_{\rm diff}\ . 
\end{eqnarray}
Here, we have used the fact that the super-Weyl weight of the massless chiral fields are $-2$
so that Eq.\,(\ref{eq:matter kin}) is invariant under the super-Weyl symmetry.
Thus, due to the Konishi-Shizuya anomaly~\cite{Konishi:1985tu}, we find that the super-Diffeomorphism 
invariant measure is not invariant under the $F$-type super-Weyl transformation.
Instead, the $F$-type super-Weyl invariant measures are given by
\begin{eqnarray}
\label{eq:SW measure}
[D Q_{\rm SW}]& \equiv& [D  \left(2 {\cal E}\right)^{1/3} Q ] = [D  \left(2 {\cal E}\right)^{-1/6} Q_{\rm diff} ]\ ,\\
\left[D \bar{Q}_{\rm SW}\right]&\equiv & [D  \left(2 {\cal E}\right)^{1/3} \bar{Q} ] = [D  \left(2 {\cal E}\right)^{-1/6} \bar{Q}_{\rm diff} ]\ ,
\end{eqnarray}
where $Q_{\rm SW}$ and $\bar{Q}_{\rm SW}$ are invariant under the the $F$-type super-Weyl transformation.
Here, the weighted chiral superfields  $Q_{\rm SW}$ and $\bar{Q}_{\rm SW}$ have  density weights $1/3$.

It should be commented that the component fields of $Q_{\rm SW}$ ($\bar{Q}_{\rm SW}$)  defined by
\begin{eqnarray}
Q_{\rm SW} &=& e^{1/3}[A_{Q_{\rm SW}} + \sqrt{2} \Theta \chi_{Q_{\rm SW}} + \Theta^2 F_{Q_{\rm SW}}]\ ,
\label{eq:SW component}
\end{eqnarray}
have the canonical kinetic terms at the leading order which decouple from the supergravity multiplets in the flat limit.
That is, for a generic chiral scalar superfield, $X = A + \sqrt{2} \Theta \chi + \Theta^2 F$, 
the chiral projection of its complex conjugate is given by
\begin{eqnarray}
\label{eq:chiral projection}
\left({\cal D}^2-8R\right)X^\dag & = & - 4 F^* + \frac{4}{3} M A^* +
\Theta^\alpha \left[
-4 i \sqrt{2} \sigma^m \partial_m \chi^\dag
\right]\nonumber\\
&&
+\Theta^2\left[
-4 \partial^2 A^*
-\frac{8}{3} M^* F^* +\frac{8}{9} A^* |M|^2 
\right] + \cdots\ ,
\end{eqnarray}
where the ellipses denote higher dimensional terms.
Then, by remembering that the component fields of $Q_{\rm SW}$ are related to those of $Q$ via
\begin{eqnarray}
Q = \left( 1 + \frac{1}{3}M^* \Theta^2\right)
\left(A_{Q_{\rm SW}} + \sqrt{2} \Theta \chi_{Q_{\rm SW}} + \Theta^2 F_{Q_{\rm SW}}\right) + \cdots\ ,
\label{eq:SW component2}
\end{eqnarray}
we find that the kinetic terms of the component fields of $Q_{SW}$ are canonical and decouple from $M$.%
\footnote{In terms of the component fields of $Q$, $M$ does not decouple 
from the kinetic term and mixes with the scalar fields via, $M^* F_Q^* A_Q$
as well as $|A_Q|^2|M|^2$ terms.}
Therefore, it is appropriate to identify the component fields of $Q_{SW}$ as the component fields 
of the corresponding chiral field in the rigid supersymmetry,%
\footnote{Here, we have neglected higher dimensional terms.
If we take them into account, we need to perform a K\"ahler-Weyl transformation to achieve 
the canonical normalisation in the Einstein frame.
}
\begin{eqnarray}
Q_{\rm rigid\,\,supersymmetry}  = A_{Q_{\rm SW}} + \sqrt{2} \theta \chi_{Q_{\rm SW}} + \theta^2 F_{Q_{\rm SW}}\ ,
\end{eqnarray}
with $\theta$ being the fermionic coordinate of the rigid superspace.

\subsection{Gaugino mass in the Wilsonian effective action}
\label{sec:mass}
As we have discussed in the previous section, 
the gaugino mass vanishes if the $F$-type super-Weyl symmetry is preserved,
and it is generated only through violations of the $F$-type super-Weyl symmetry.
As relevant terms of the gauge supermultiplet
preserve the super-Weyl symmetry, the gaugino mass appearing in the super-Diffeomorphism 
invariant ``tree-level" Wilsonian action is highly suppressed.

The approximate $F$-type super-Weyl symmetry is, however, 
anomalously broken by the super-Diffeomorphism invariant measure $[D Q_{\rm diff}]$.
To read off the gaugino mass from this violation, it is transparent to change 
the path-integral measure to the $F$-type super-Weyl invariant measure, $[D Q_{\rm SW}]$,
so that the super-Weyl variance is apparent in the corrected ``tree-level" Wilsonian action. 
In fact, the change of the measures from $[DQ_{\rm diff}]$ to $[DQ_{SW}]$
is accompanied by the Konishi-Shizuya anomaly~\cite{Konishi:1985tu},%
\footnote{The identity in Eq.~(\ref{eq:translation}) is not quite correct.
In general, ${\mit \Delta} S$ involves higher dimensional terms suppressed by the cut off of the Wilsonian effective action.
However, such higher-dimensional terms are negligible.}
\begin{eqnarray}
\label{eq:translation}
[D Q_{\rm diff}] [D\bar{Q}_{\rm diff}] [D Q^\dag_{\rm diff}] [D\bar{Q}^\dag_{\rm diff}] &=&
[D Q_{\rm SW}] [D\bar{Q}_{\rm SW}] [D Q^\dag_{\rm SW}] [D\bar{Q}^\dag_{\rm SW}]
\times {\rm exp}
\left[
i {\mit\Delta} S
\right],\nonumber\\
{\mit\Delta} S &=& \frac{1}{16} \frac{1}{2\pi^2}\times \int {\rm d}^4x\, {\rm d}^2\Theta\, 2{\cal E}\, {\ln}(2{\cal E})^{1/6}\, W^\alpha W_\alpha + {\rm h.c.}\ .
\end{eqnarray}
Accordingly, the ``tree-level" Wilsonian effective action which should be taken in conjunction with  $[D Q_{\rm SW}]$
is given by,
\begin{eqnarray}
\label{eq:SSW}
S = S_{SD} + {\mit \Delta}S\ .
\end{eqnarray}
Here, $S_{SD}$ denotes the super-Diffeomorphism invariant local Wilsonian effective action discussed above.
Without surprise, ${\mit\Delta} S$ is not invariant under the super-Diffeomorphism, 
which cancels the anomalous breaking of the super-Diffeomorphism invariance 
by $[D Q_{\rm SW}]$.
We summarize properties of the measures in Table.~\ref{tab:measures}.%
\footnote{Throughout this paper, we presume the regularization scheme of the path-integral measure
which reproduce the  the Konishi-Shizuya anomaly in the form in Eq.\,(\ref{eq:translation}).
In the dimensional regularization/reduction, on the other hand, 
the change of the path-integral measures is not accompanied by the rescaling anomaly,
while the approximate super-Weyl symmetry is explicitly broken by the relevant interactions 
which eventually leads to a consistent gaugino mass\,\cite{Boyda:2001nh}.
}
\begin{table}[tb]
\begin{center}
\begin{tabular}{c||c|c|c|}
    & measure & action & gaugino mass\\ \hline
\,$[D Q_{\rm diff}]$ \,&\, $SD$,\,  $\cancel{SW}$ \, & \,$SD$,\, $SW$ \,& \, hidden in the measure\, \\
\,$[D Q_{\rm SW}]$ \,& \,$\cancel{SD}$,\, $SW$ \, & $\,\cancel{SD}$,\, $\cancel{SW}$\,  &\, apparent in the action\,
\end{tabular}
\end{center}
\caption{Properties of two path-integral measures.
Here, $SD$ and $SW$ denote the super-Diffeomorphism and the  
$F$-type super-Weyl invariances, respectively.
The cancel lines denote non-invariances.
}
\label{tab:measures}
\end{table}%

Armed with a correct ``tree-level" Wilsonian action along with the super-Weyl invariant measure, 
we can now read off the gaugino mass directly from the local term in the action, ${\mit \Delta}S$,
which leads to
\begin{eqnarray}
\label{eq:gaugino mass}
m_\lambda/g^2= - \frac{1}{2} \frac{1}{2\pi^2} {\ln}(2{\cal E})^{1/6}|_{\Theta^2} =  \frac{1}{24\pi^2} M^* = -\frac{1}{16\pi^2}\times 2m_{3/2}\ ,
\end{eqnarray}
where ${\cal X}|_{\Theta^2}$ denotes the $\Theta^2$ component of a superfield ${\cal X}$.
This gaugino mass reproduces the anomaly mediated gaugino mass given in~\cite{Randall:1998uk,Giudice:1998xp}.
In this way, we find that the anomaly mediated gaugino mass can be read off from 
the super-Diffeomorphism non-invariant term ${\mit\Delta} S$ in the superspace formalism of supergravity.%
\footnote{In this paper, we concentrate on the anomaly mediated gaugino mass at one-loop level.}

\subsection{Radiative corrections from path-integration}
So far, we have fixed the Wilsonian scale to $M_{\rm PL}$ and have not performed any path-integration. 
Here, let us discuss effects of the path-integration.
After integrating out modes above a scale $\Lambda (<M_{\rm PL})$,
the Wilsonian effective action at $\Lambda$ is again given by the form of Eq.\,(\ref{eq:SSW}),
with renormalized coefficients and higher dimensional operators suppressed not only by $\Mpl$ but also by $\Lambda$.
Due to the presence of cutoff scales, the super-Weyl symmetry in the Wilsonian action at the scale $\Lambda$
is hardly preserved.
As we have discussed, however, the relevant terms of the matter and the gauge supermultiplets 
have an approximate super-Weyl symmetry accidentally due to the super-Diffeomorphism invariance.
Therefore, radiative corrections
do not generate the gaugino mass term 
beyond the one in Eq.~(\ref{eq:gaugino mass}) up to $\Lambda$ or $\Mpl$ suppressed corrections.

It should be also noted that, among various corrections, the ones 
from diagrams which involve  Planck suppressed interactions
lead to higher dimensional operators suppressed at least by a single power of $M_{\rm PL}$ in the effective action 
at $\Lambda$.%
\footnote{
If there are ultraviolet divergences which are cancelled only by non-local terms, 
$M_{\rm PL}$ suppressed interactions could lead to higher dimensional operators suppressed not 
by $M_{\rm PL}$ but only by $\Lambda$ at the cutoff scale $\Lambda$.
The Bogoliubov-Parasiuk-Hepp-Zimmermann prescription~\cite{Bogoliubov:1957gp,Hepp:1966eg,Zimmermann:1969jj} shows that ultraviolet divergences in general can be renormalized away by local terms.
}
Effects to lower dimensional operators through ultra-violet divergences are renormalized 
by the shifts of the corresponding operators~\cite{Polchinski:1983gv}.
Visible effects of higher dimensional operators only show up 
through higher dimensional operators even in the effective action at $\Lambda$.

Concretely, radiative corrections from loop diagrams involving gravity supermultiplets (in particular gravitinos with small momenta)
may lead to higher dimensional operators such as $|M|^n M^* \lambda\lambda$ ($n\ge 0$)
suppressed only by $M_{\rm PL}^2 \Lambda^{n-2}$.
Such diagrams involving the gravitinos however damp for $\Lambda \ll m_{3/2}$.
Therefore, they contribute to the gaugino mass at most of $O(m_{3/2}^3/M_{\rm PL}^2)$.

From these arguments, we see that higher dimensional operators which are suppressed by not $M_{\rm PL}$ 
but only by $\Lambda$ in the Wilsonian effective action at the cutoff scale $\Lambda$ 
are generated only from relevant interactions of the matter and gauge supermultiplets.
Such effects can be properly taken care of within the renormaizable effective theory
of the matter and the gauge supermultiplets  with softly broken supersymmetry.

Let us emphasize again that the super-Diffeomorphism violation is not arbitrary in the Wilsonian effective action at $\Lambda$, although the super-Diffeomorphism invariance is broken by $[DQ_{\rm SW}]$.
The super-Diffeomorphism violation in the Wilsonian action is uniquely given  by ${\mit\Delta S}$ 
at each Wilsonian scale, so that the super-Diffeomorphism is preserved in the quantum theory.
Thus, the accidental approximate super-Weyl symmetry which is the outcome of the super-Diffeomorphism
invariance is justified even after performing path-integration.

Putting all together, we  find that the anomaly mediated gaugino mass can be extracted
from the super-Diffeomorphism non-invariant local term in the Wilsonian effective action
at the scale $\Lambda \gg m_{3/2}$ in the superspace formalism of the supergravity.
Radiative corrections to the gaugino mass operator are dominantly given by relevant interactions 
of the matter and the gauge supermultiplets.
Therefore, once we extract a gaugino mass at some high cutoff scale, we can use the gaugino mass
as the boundary condition of the renormalization group equation at $\Lambda$ 
in the low-energy effective renormalizable supersymmetric theory with soft supersymmetry breaking.

\subsection{Decoupling effects of massive matter}
Before closing this section, let us consider the contribution to the gaugino mass 
from charged matter multiplets with a supersymmetric mass $m$ far larger than $m_{3/2}$,
\begin{eqnarray}
{\cal L}_{\rm mass } = \int {\rm d}^2\Theta\, 2{\cal E} \,m Q \bar{Q} + {\rm h.c.} \ .
\end{eqnarray}
If the cutoff scale of the Wilsonian effective action is far above $m$, the mass $m$ is negligible 
in comparison with the kinetic term and hence the above discussion holds. 
When the cutoff scale is below $m$, the mass term dominates over the kinetic term.
In that situation, the approximate super-Weyl symmetry is such that the mass term is invariant.%
\footnote{In the Pauli-Villars regularization, the anomaly mediated gaugino mass is understood by the difference of super-Weyl invariant measures between massive Pauli-Villars fields and massless matter fields (see Appendix~\ref{sec:Pauli-Villars}).}
This observation leads to the Weyl weights of $-3$ for $Q$ and $\bar{Q}$, i.e. 
$\delta _{\rm SW,massive}Q = -3 \Sigma Q + \cdots$, and hence,
the super-Weyl invariant measures of the massive matter are given by
\begin{eqnarray}
\label{eq:SW mass measure}
[D Q_{\rm SW,massive}] \equiv [D  \left(2 {\cal E}\right)^{1/2} Q ]\ , 
\quad
[D \bar{Q}_{\rm SW,massive}] \equiv [D  \left(2 {\cal E}\right)^{1/2} \bar{Q} ]\ ,
\end{eqnarray}
which coincide with the super-Diffeomorphism invariant measures in Eq.\,(\ref{eq:SD measure}).
Thus, below the scale $m$, the approximate super-Weyl symmetry is well described by 
the super-Diffeomorphism invariant Wilsonian effective action, i.e. ${\mit \Delta} S = 0$, 
and hence, no anomaly mediated gaugino mass term appears up to $O(m_{3/2}^2/m)$ contributions.
This argument reconfirms the insensitivity of the anomaly mediated gaugino mass to ultraviolet physics~\cite{Giudice:1998xp}.

If $m$ is close to $m_{3/2}$, the decoupling does not hold in general.
The Wilsonian effective action below the mass threshold of $Q$ and $\bar{Q}$ includes terms suppressed only by $m$, which might contribute to the gaugino mass as large as $m_{3/2}^2/m$.
Integration of $Q$ and $\bar{Q}$ should be performed explicitly, as is the case with the higgsino threshold correction in the minimal supersymmetric standard model~\cite{Giudice:1998xp}.

\section{Fictitious super-Weyl gauge symmetric formulation}
\label{sec:fictitious SW}
In the discussion in \cite{deAlwis:2008aq,D'Eramo:2013mya}, 
the origin of the gaugino mass has been discussed in the superspace formalism of the supergravity
with the help of a fictitious (and exact) super-Weyl gauge symmetry 
by introducing a chiral super-Weyl compensator field, $C$, in the track of \cite{Kaplunovsky:1994fg}. 
We call this super-Weyl symmetry as the fictitious super-Weyl gauge symmetry throughout the paper
to distinguish it from the approximate super-Weyl symmetry we have discussed so far.
One of the key to settle the puzzle in the discussion  in \cite{deAlwis:2008aq,D'Eramo:2013mya}
is how to write down the anomaly mediated gaugino mass term
in a gauge independent way of the fictitious super-Weyl gauge symmetry.
In this section, we show how to write down the gauge independent gaugino mass term,
where the knowledge on the super-Diffeomorphism invariant path-integral measure plays a crucial role.

\subsection{Fictitious super-Weyl gauge symmetry}
The fictitious (and exact) super-Weyl gauge symmetry is introduced to the action in Eq.~(\ref{eq:classical action}) 
by performing a finite super-Weyl transformation in Eq.~(\ref{eq:SW}) with $\Sigma = {\ln}\,C/2$ and $w=0$~\cite{Kaplunovsky:1994fg}.
The resulting classical acton is given by
\begin{eqnarray}
\label{eq:SW classical action}
{\cal L} &=&  \Mpl^2 \int {\rm d}^2 \Theta\,2 {\cal E}'  \frac{3}{8} \left({\cal D}'^{\dag2} - 8R'\right)C C^\dag{\exp}\left[-\frac{K'}
{3\Mpl^2}\right] \nonumber\\
&& + 
\frac{1}{16g^2}\int {\rm d}^2 \Theta\, 2{\cal E}'\, W'^\alpha W'_\alpha + {\rm h.c.}\ ,
\end{eqnarray}
where primes denote fields after the transformation.
Now, the action is exactly invariant under the super-Weyl symmetry in Eq.\,(\ref{eq:SW})
in terms of ${\cal E}'$, $W'^\alpha$, $Q'$ and $\bar{Q}'$ with $w = 0$,
while giving a Weyl weight $-2$ to the ``super-Weyl compensator" $C$, 
\begin{eqnarray}
\delta_{\rm SW,fic} C &=& -2 \Sigma C -S^\alpha \frac{\partial}{\partial \Theta^\alpha}C\ .
\end{eqnarray}
It should be noted that the compensator $C$ is a gauge degree of freedom, which can be completely eliminated 
by performing the fictitious super-Weyl transformation.
In other words, one may take any $C$ so that a calculation one performs is as simple as possible.%
\footnote{The singular transformation leading to $C = 0$ should be avoided.}
In particular, in the presence of the compensator, the equation of the motion of $M'$ is changed from Eq.\,(\ref{eq:auxM})
to 
\begin{eqnarray}
F^C - \frac{1}{3}M'^* = m_{3/2} \ ,
\end{eqnarray}
where we have taken $C = 1 + F^C \Theta^2$.
Thus, for example, it is convenient to take the gauge where $M' = 0$, which is taken 
in \cite{D'Eramo:2013mya} up to higher dimensional terms (see also \cite{Cheung:2011jp}).

\subsection{Gaugino mass}
As we have discussed in the previous section, the super-Weyl transformation performed to introduce $C$ is anomalous
where the measure is transformed from $[D Q_{\rm dff}]$ to $[D Q'_{\rm dff}]$.%
\footnote{The weighted chiral field $Q_{\rm diff}$ has a Weyl weight $3$ for $w = 0$.}
The transformation invokes the following term in the Wilsonian effective action,
\begin{eqnarray}
\label{eq:counter' SW}
\Delta S'_C =  + \frac{1}{16} \frac{3}{4\pi^2} \int {\rm d}^4 x\, {\rm d}^2\Theta \,2{\cal E}'\, {\ln}\,C \,W'^\alpha W'_\alpha + {\rm h.c.}\ .
\end{eqnarray}
This term can be also derived from the condition that the fictitious super-Weyl symmetry is free from 
the gauge anomaly~\cite{Kaplunovsky:1994fg}.
Further, let us eliminate $C$ from the kinetic term of the matter fields by the redefinitions, $Q'' \equiv Q'C$ and $\bar{Q}''\equiv \bar{Q}'C$.
After the redefinitions, the integration of the matter fields does not generate the gaugino mass proportional to $F^C$
at one-loop level,
so that the gaugino mass is directly read off from the Wilsonian effective action.
By combining the counter terms of the anomalies to reach to 
$Q''_{\rm diff} = (2 {\cal E'})^{1/2} Q'C$ 
and 
$\bar{Q}''_{\rm diff} = (2 {\cal E'})^{1/2} \bar{Q}'C$, we eventually obtain
\begin{eqnarray}
\label{eq:counter SW}
\Delta S_C =   \frac{1}{16} \frac{1}{4\pi^2}\times \int {\rm d}^4 x\, {\rm d}^2\Theta\, 2{\cal E}'\,{\ln}\,C \,W'^\alpha W'_\alpha + {\rm h.c.}\ ,
\end{eqnarray}
where the corresponding path-integral measures are given by $[D Q''_{\rm dff}]$ and $[D \bar{Q}''_{\rm dff}]$.

In \cite{deAlwis:2008aq}, 
it is claimed that there is no anomaly mediated gaugino mass derived in \cite{Randall:1998uk,Giudice:1998xp} 
by taking a gauge with $F^C = 0$. 
On the other hand, in \cite{D'Eramo:2013mya}, taking another gauge with $M'=0$, the anomaly mediated gaugino mass is reproduced.
These arguments pose a puzzle, for the gaugino mass  should not depend on the gauge choice of $F^C$.

This puzzle is solved by remembering the discussion in Sec.~\ref{sec:measure}.
There, in order to read off the gaugino mass from the Wilsonian effective action, we have used the canonical measure $[D Q_{\rm SW}]\equiv [D  \left(2 {\cal E}\right)^{1/3} Q ]$.
Similarly, we should use again the measure,
\begin{eqnarray}
[DQ_c] \equiv [D  \left( 2 {\cal E'}\right)^{1/3} C Q' ] = [D  \left( 2 {\cal E'}\right)^{-1/6} Q''_{\rm diff}  ]\ ,
\end{eqnarray}
which is again invariant under the ``approximate" super-Weyl symmetry.
The kinetic term of $Q_c$ is free from the mixings to both $M'$ and $F^C$, and hence, canonical.
Eventually, by translating the measure from $[DQ_{\rm diff}'']$ to $[DQ_{c}]$, the Wilsonian 
effective action obtains a correction ${\mit\Delta} S$, which add up with ${\mit\Delta} S_C$,%
\footnote{One may obtain the following counter term directly from the relation,
\begin{eqnarray}
[DQ_{c}] = [D (2{\cal E})^{-1/6} C^{-1/2} Q_{\rm diff}]\ .
\end{eqnarray}
}
\begin{eqnarray}
\label{eq:SWinvariantForm}
{\mit\Delta} S + {\mit\Delta} S_C =   \frac{1}{16} \frac{1}{4\pi^2}\times \int {\rm d}^4 x\, {\rm d}^2\Theta \,2{\cal E}' 
\left({\ln}\left(2 {\cal E}'\right)^{1/3} + {\ln}\,C \right)W'^\alpha W'_\alpha + {\rm h.c.}\ .
\end{eqnarray}
This expression is manifestly invariant under the fictitious super-Weyl transformation.
Again the counter term is not invariant under the super-Diffeomorphism, which is inevitable 
to cancel the anomaly of the super-Diffeomorphism due to  $[DQ_c]$.
From this expression, we obtain the anomaly mediated gaugino mass 
\begin{eqnarray}
 m_\lambda/g^2 = - \frac{1}{2} \frac{1}{4\pi^2}
\left( {\ln}(2{\cal E}')^{1/3}+{\ln}\,C\right)|_{\Theta^2} = - \frac{1}{8\pi^2} \left( F^C - \frac{1}{3} M'^*\right) =- \frac{1}{16\pi^2}\times 2m_{3/2}\ ,
\label{eq:SWinvGM}
\end{eqnarray}
which is independent of the gauge choice of $F^C$.

In our argument, the super-Diffeomorphism variant counter term ${\mit \Delta }S$ is the key
to obtain the manifestly invariant expression of the anomaly mediated gaugino mass 
under the fictitious super-Weyl gauge symmetry.
It should be also stressed that the combination,
\begin{eqnarray}
 \int {\rm d}^4 x\, {\rm d}^2\Theta \,2{\cal E}' 
\left({\ln}\left(2 {\cal E}'\right)^{1/3} + {\ln}\,C \right)W'^\alpha W'_\alpha + {\rm h.c.}\ ,
\label{eq:invariant combination}
\end{eqnarray}
is invariant under the fictitious super-Weyl symmetry.
Thus, the mere knowledge of the anomaly of the fictitious super-Weyl gauge symmetry 
cannot determine the overall coefficient of Eq.~(\ref{eq:SWinvariantForm}), and it is crucial to start with the super-Diffeomorphism invariant measure 
to obtain Eq.\,(\ref{eq:SWinvariantForm}).%
\footnote{Correspondingly, in the 1PI effective action, the fictitious super-Weyl gauge invariance alone cannot determine the gaugino mass term up to the contribution from Eq.~(\ref{eq:invariant combination}) with ${\ln}({2\cal E'})^{1/3}$ replaced by ${\rm ln} \,\Omega^{-1}$, where the chiral field $\Omega$ is defined in the following.
}

\subsection{Relation to the 1PI quantum effective action (I)}
\label{sec:1PI}
As is clear from Eq.\,(\ref{eq:SWinvGM}), the gaugino mass is simply read off from 
the counter term in the Wilsonian effective action, ${\mit\Delta} S_C$, 
by taking the gauge with $M'=0$ and $F^C = m_{3/2}$.
In the 1PI quantum effective action, on the other hand, it should be also possible 
to write down the gaugino mass term without using the compensator $C$.
To see how the gaugino mass appear in the 1PI action, let us 
consider a finite super-Weyl transformation of $R$,
\begin{eqnarray}
R' = - \frac{1}{8}e^{4 \Sigma} \left( {\cal D}^{\dag 2} - 8 R \right) e^{-2 \Sigma^\dag} + \cdots\ .
\end{eqnarray}
Here, ellipses denote terms which are irrelevant for the transformation of the lowest component of $R$.
Then, by taking $\Sigma$ such that 
\begin{eqnarray}
\label{eq:eliminate R}
\left( {\cal D}^2 - 8 R^\dag \right) e^{-2 \Sigma} =0\ ,
\end{eqnarray}
we can eliminate the lowest component of $R$.
The solution of Eq.\,(\ref{eq:eliminate R}) is given by\,\cite{Gates:1983nr,Butter:2013ura};
\begin{eqnarray}
e^{-2\Sigma} \equiv \Omega = 1 + \frac{1}{2\Box_+} \left({\cal D}^{\dag 2} - 8R\right) R^\dag\ , \nonumber \\
\Box _+ \equiv  \frac{1}{16} \left({\cal D}^{\dag 2} - 8R\right) \left({\cal D}^2 - 8R^\dag\right)\ .
\end{eqnarray}
Thus, by setting $C= \Omega^{-1}$, 
we can achieve the desirable gauge choice of the fictitious super-Weyl gauge symmetry where $M'=0$.
It should be noted that the apparent non-local expression of $\Omega$ does not cause problems since
the chiral field $\Omega$ is reduced to a local field expression,
\begin{eqnarray}
\Omega \simeq 1 + \frac{1}{3} M^* \Theta^2\ ,
\label{eq:Omega}
\end{eqnarray}
in the flat limit. 
Thus, as long as we are interested in the flat limit, $\Omega$ can be treated as a local field.

In this gauge, $\Delta S_C$ is now expressed by,
\begin{eqnarray}
{\mit\Delta} S_{C=\Omega^{-1}} =  \frac{1}{16} \frac{1}{4\pi^2}\times \int {\rm d}^4x\, {\rm d}^2\Theta \,2{\cal E}' \,{\ln}\,\Omega^{-1}\, W'^\alpha W'_\alpha + {\rm h.c.}\ .
\end{eqnarray}
By expanding this expression around $\Omega =1$, we obtain 
\begin{eqnarray}
\label{eq:non-local}
{\mit\Delta} S_{C = \Omega^{-1}} \simeq -  \frac{1}{16} \frac{1}{8\pi^2}  \int {\rm d}^4x\, {\rm d}^2\Theta\, 2{\cal E} \,\frac{1}{\Box_+} \left({\cal D}^{\dag 2} - 8R\right) R^\dag \, W^\alpha W_\alpha + {\rm h.c.} \ ,
\end{eqnarray}
at the leading order.
Here, we have reverted ${\cal E}'$ and $W^{\prime\alpha}$ to 
 ${\cal E}$ and $W^\alpha$.
Since this term is expressed in terms of the gravity multiplet and independent of $C$, this provides 
an appropriate expression of the super-Weyl variance in the 1PI effective action.
In fact, the final expression reproduces the 1PI quantum effective action given in \cite{Bagger:1999rd}.%
\footnote{Apparent difference by a factor of $4$ between our result and that in \cite{Bagger:1999rd} is due to the difference of the normalization of the gauge multiplet.}
By substituting Eq.\,(\ref{eq:Omega}), 
we again obtain the anomaly mediated gaugino mass in~\cite{Randall:1998uk,Giudice:1998xp}.

%%%%%%%%%%%%%%%%%%%%%%%%%%%%%%%
\subsection{Relation with 1PI quantum effective action (II)}
The chiral field $\Omega$ is also useful to discuss the 1PI quantum effective action along the 
lines of Sec.\,\ref{sec:measure}, where we have not introduced the super-Weyl compensator $C$.
There, instead, we relied on the $F$-type super-Weyl invariant but super-Diffeomorphism variant measure
to read off the gaugino mass from the Wilsonian effective action. 
The 1PI quantum effective action, however, must be invariant under the super-Diffeomorphism by itself.
Thus, ${\mit \Delta} S$ should be replaced by a super-Diffeomorphism invariant expression
in the 1PI quantum effective action.

To find an appropriate expression, let us remember that the chiral field $\Omega$ transforms,
\begin{eqnarray}
\delta _{\rm SW} \Omega =  -2\Sigma \Omega -  S^\alpha \frac{\partial}{\partial \Theta^\alpha}\Omega\ ,
\end{eqnarray}
under the super-Weyl transformation.
From this property, we can construct a measure
\begin{eqnarray}
[D Q_{\rm SW,diff}] \equiv [D \Omega^{1/2} \left(2 {\cal E}\right)^{1/2}Q ] = [D \Omega^{1/2} Q_{\rm diff} ]\ ,
\end{eqnarray}
which is invariant under both the $F$-type super-Weyl and the super-Diffeomorphism transformations.%
\footnote{The component fields of $Q_{\rm SW,diff}$ defined by,
$Q_{\rm SW,diff}=e^{1/2} [A_{Q_{\rm SW,diff}} + \sqrt{2} \Theta \chi_{Q_{\rm SW,diff}} + \Theta^2 F_{Q_{\rm SW,diff}}] $,
have the same canonical kinetic term with those of $Q_{SW}$ in Eq.\,(\ref{eq:SW component}).
}
Thus, in a similar way as Sec.\,\ref{sec:measure},
the Wilsonian effective action receives  a correction by 
changing the measure from $[DQ_{\rm diff}]$ to $[DQ_{\rm SW,diff}]$, 
\begin{eqnarray}
\label{eq:translation2}
[D Q_{\rm diff}] [D\bar{Q}_{\rm diff}] [D Q^\dag_{\rm diff}] [D\bar{Q}^\dag_{\rm diff}] &=&
[D Q_{\rm SW,diff}] [D\bar{Q}_{\rm SW,diff}] [D Q^\dag_{\rm SW,diff}] [D\bar{Q}^\dag_{\rm SW,diff}]
\times {\rm exp}
\left[
i {\mit\Delta} S_{\rm diff}
\right],\nonumber\\
{\mit\Delta} S_{\rm diff}&=&  \frac{1}{16} \frac{1}{4\pi^2}\times \int {\rm d}^4x\, {\rm d}^2\Theta\, 2{\cal E} \,{\ln}\,\Omega^{-1}\, W^\alpha W_\alpha + {\rm h.c.}\ .
\end{eqnarray}
Unlike ${\mit \Delta} S$, ${\mit\Delta} S_{\rm diff}$ is invariant under the super-Diffeomorphism.
Thus, ${\mit\Delta} S_{\rm diff}$ is an appropriate expression of the super-Weyl breaking in the 
1PI quantum effective action.
Again, this expression reproduces the super-Weyl breaking term in the 1PI effective action in \cite{Bagger:1999rd}.

\section{Summary}
\label{sec:summary}
In this paper, we have reexamined the anomaly mediated gaugino mass term in the superspace formalism of supergravity.
The absence of the gaugino mass term of $O(m_{3/2})$ in the classical supergravity action is understood 
by an approximate super-Weyl symmetry of the super-Diffeomorphism invariant local classical action.
Then, we find that the anomaly mediated gaugino mass originates from the anomalous breaking of 
the approximate super-Weyl symmetry caused by the super-Diffeomorphism invariant measure 
of the charged field.
By changing the path-integral measure from the super-Diffemorphism invariant one 
to the super-Weyl invariant one, we have shown that the gaugino mass term can be 
read off from the local counter term in the Wilsonian action.
It should be stressed that the counter term is not invariant under the super-Diffeomorphism, which 
is required for the super-Diffeomorphism invariance of the quantum theory.
As is clear from our discussion, the path-integral measure plays a crucial role in determining the gaugino mass term.
This observation fills a gap in the literature on the anomaly mediated gaugino mass.

We have also discussed the gaugino mass in the formulation with a fictitious super-Weyl gauge symmetry.
There, the action is made invariant under a fictitious super-Weyl gauge 
symmetry by introducing a chiral compensator $C$.
Since $C$ is a gauge degree of freedom, the gaugino mass should be independent of the choice of the value of $C$.
A gauge independent expression of the local gaugino mass term was, however,
not known in the literature, which is one of the origins of the controversy in \cite{deAlwis:2008aq,D'Eramo:2013mya}.
In our discussion, we have shown how the gauge independent expression is obtained 
with the aid of the super-Diffeomorphism invariant measure.
We have also discussed how to reconcile the gaugino mass term appearing in 
the non-local 1PI effective action given in \cite{Bagger:1999rd} and the one in the local effective Wilsonian action.

In our discussion, we have concentrated on the anomaly mediated gaugino mass at one-loop level.
At one-loop level,
the violation of the $F$-type super-Weyl symmetry in the gauge kinetic function, which is the origin of the gaugino mass, is 
extracted by calculating the anomalous Jacobian associated with the change of the measure from the super-Diffeomorphism invariant one to the $F$-type super-Weyl invariant one (see Eq.~(\ref{eq:translation})).
This corresponds to the fact that a one-loop beta function of a gauge theory
is extracted by calculating the anomalous Jacobian associated with the Weyl transformation~\cite{Fujikawa:1993xv}.
Note that the approximate super-Weyl symmetry is also broken by anomalous dimensions of the matter and gauge multiplets,
which contributes to the gaugino mass at two and more loop level.
These contributions are difficult to be extracted in the Wilsonian effective action.
It would be easier to discuss the violation of the super-Weyl symmetry in the 1PI effective action.

As a final remark, let us sketch the gaugino mass in a non-Abelian gauge theory.
In the non-Abelian gauge theory, the path-integral measure of the gauge multiplet should be taken into account.
The super-Diffeomorphism invariant measure and the $F$-type super-Weyl invariant measure are given by
\begin{eqnarray}
[DV_{\rm diff}] = [D E^{1/2} V],~~
[DV_{\rm SW}] =  [D \left(2 {\cal E}\right)^{-1/6} \left(2 {\cal E}^\dag\right)^{-1/6} V_{\rm diff} ]\ ,
\end{eqnarray}
where $E$ is the determinant of the supersymmetric vielbein in a real superspace.
The super-Weyl transformation law of $E$ is given by
\begin{eqnarray}
\label{eq:gauge measures}
\delta_{\rm SW} E = 2 \left(\Sigma + \Sigma^\dag \right) E + \cdots\ ,
\end{eqnarray}
where the ellipses denote inhomogeneous terms which can be cancelled by the super-Diffeomorphism.
Here, we collectively represents the gauge multiplet 
and the ghost multiplets by $V$, and so are $V_{\rm diff}$ and $V_{\rm SW}$ accordingly.

The translation from $[D V_{\rm diff}] $ to $[D V_{\rm SW}]$ is easily performed in the following way.
Let us introduce a chiral compensator $C$ as in Sec.~\ref{sec:fictitious SW},
which defines $E'$ via,
\begin{eqnarray}
\label{eq:E SW}
E = C C^\dag E'\ .
\end{eqnarray}
By remembering that the super-Weyl transformation is anomalous, the gauge kinetic function receives 
a counter term depending on $C$ as \cite{Kaplunovsky:1994fg}\ ,
\begin{eqnarray}
\label{eq:gauge anomaly}
[D E^{1/2} V] &=& [D E'^{1/2} V ] \times e^{i {\mit\Delta} S^V_C}\ ,\nonumber\\
{\mit\Delta }S^V_C &=&  - \frac{1}{16} \frac{3T_G}{8\pi^2} \times\int {\rm d}^4 x\, {\rm d}^2\Theta \,2{\cal E}' \,{\ln}\,C\, W'^\alpha W'_\alpha + {\rm h.c.}\ ,
\end{eqnarray}
where $T_G$ is the Dynkin index of the adjoint representation.
It should be noted that ${\mit\Delta} S^V_C$ includes the rescaling anomaly form the ghost multiplets.
Then, by comparing Eqs.~(\ref{eq:gauge measures}), (\ref{eq:E SW}) and (\ref{eq:gauge anomaly}), 
we find that the counter term appearing in translation from $[D V_{\rm diff}] $ to $[D V_{\rm SW}]$ 
is given by replacing $C$ to $(2{\cal E})^{1/3}$,%
\footnote{The expression of the rescaling anomaly does not depend on whether 
the rescaling factor is a chiral superfield or a chiral density superfield.}
 which leads to
\begin{eqnarray}
\label{eq:}
\label{eq:translation2}
[D V_{\rm diff}] = [D V_{\rm SW} ] \times e^{i \Delta S^V_C}\ ,~~ C = \left(2 {\cal E}\right)^{1/3}\ .
\end{eqnarray}

By putting Eqs.~(\ref{eq:translation}) and (\ref{eq:translation2}) together, we obtain
\begin{eqnarray}
\prod_R[DQ^R_{\rm diff}] [DQ^{R\dag}_{\rm diff}] [DV_{\rm diff} ] = \prod_i[DQ^R_{\rm SW}] [DQ^{R\dag}_{\rm SW}] [DV_{\rm SW} ] \times e^{i{\mit \Delta} S}\ ,\nonumber \\
{\mit\Delta} S = - \frac{1}{16} \frac{3T_G -\sum T_R}{8\pi^2}\times \int {\rm d}^4 x\, {\rm d}^2\Theta\, 2{\cal E} \,{\ln}\left(2 {\cal E} \right)^{1/3} W^\alpha W_\alpha + {\rm h.c.}\ ,
\end{eqnarray}
where $T_R$ is the total Dynkin index of matter fields $Q^R$. 
As a result, we find an expression of the gaugino mass,
\begin{eqnarray}
\label{eq:gaugino mass NA}
 m_\lambda/g^2 = \frac{1}{2} \frac{3 T_G - T_R}{8\pi^2} {\ln}(2{\cal E})^{1/3}|_{\Theta^2} =  \frac{3T_G - T_R}{16\pi^2}\times m_{3/2}\ ,
\end{eqnarray}
which reproduces the anomaly mediated gaugino mass found in~\cite{Randall:1998uk,Giudice:1998xp}.
We may also obtain the manifestly gauge independent expression in the fictitious super-Weyl symmetry
for the non-abelian gauge theory by using Eq.\,(\ref{eq:translation2}) along the lines of Sec.\,\ref{sec:fictitious SW}.

%---------------SECTION------------------%
%
\section*{Acknowledgements}
We thank Tsutomu T. Yanagida for useful discussion.
This work is supported by
Grant-in-Aid for Scientific research 
from the Ministry of Education, Science, Sports, and Culture (MEXT), Japan, No. 24740151 
and 25105011 (M.I.), from the Japan Society for the Promotion of Science (JSPS), No. 26287039 (M.I.),
the World Premier International Research Center Initiative (WPI Initiative), MEXT, Japan (K.H. and M.I.)
and a JSPS Research Fellowships for Young Scientists (K.H.).
The work of M.I. was supported in part by National Science Foundation Grant No. PHYS-1066293 and the hospitality of the Aspen Center for Physics.

%
%---------------SECTION------------------%

\appendix

\section{Super-Diffeomorphism invariant measure}
\label{sec:Sdiff measure}
In this appendix, we show that the measure given in Eq.~(\ref{eq:SD measure}) 
is invariant under the super-Diffeomorphism.
Under the tranformation given in Eq.~(\ref{eq:Sdiff}), the variable $Q_{\rm diff}$ transforms as
\begin{eqnarray}
Q_{\rm diff} \rightarrow Q_{\rm diff}' &=& Q_{\rm diff}- \eta^M(x,\Theta)\partial _M Q_{\rm diff} - \frac{1}{2}(-)^M \left(\partial_M \eta^M\left(x, \Theta\right)\right)Q_{\rm diff}\ .
\label{eq:QdiffSD}
\end{eqnarray}
Then, the path-integral measure $[DQ_{\rm diff}]$ transforms,
\begin{eqnarray}
[DQ_{\rm diff}'] &=& [DQ_{\rm diff}]\times {\rm exp}\left[\,{\rm sTr}\,{\cal O}(z',z)\,\right]\ ,\\
{\cal O}(z',z) &\equiv& - \left[ \eta^M\partial_M + \frac{1}{2} (-1)^M \left(\partial_M\eta^M\right) \right]\delta^6(z'-z)\ ,
\label{eq:Ozz}
\end{eqnarray}
where we have collectively represented $x$ and $\Theta$ by $z$.
Formally, the super-trace sTr is expressed by
\begin{eqnarray}
{\rm sTr}\,{\cal O}(z',z) = \int {\rm d}^6z {\rm d}^6z' \delta^6 (z'-z) {\cal O}(z',z)\ .
\label{eq:sTr}
\end{eqnarray}
A naive conclusion is that the super-trace vanishes due to the saturation of Grassmann 
variables $\Theta$ and $\Theta'$ from the delta functions in Eqs.\,(\ref{eq:Ozz}) and (\ref{eq:sTr}).
However, since there is also a factor of $\delta^4(x'-x)$, which is well-defined only after integrating over $x$ or $x'$, 
one should carefully investigate the integration.

To examine the the integration, let us expand the delta function by plane waves,
\begin{eqnarray}
\delta^6 (z'-z) &=& \int \frac{{\rm d}^4 k}{(2\pi)^4} d^2 \tau \Psi_{-k,-\tau}(z') \Psi_{k,\tau}(z)\ ,\\
\Psi_{k,\tau}(z) &\equiv& {\rm exp}(ikx+ 2i \tau \Theta)\ .
\end{eqnarray}
By substituting this expression into Eq.\,(\ref{eq:sTr}), the above super-trace is expressed by,
\begin{eqnarray}
{\rm sTr}\,{\cal O}(z',z) = - \int {\rm d}^6z \int \frac{{\rm d}^4 k}{(2\pi)^4} d^2 \tau \Psi_{-k,-\tau}(z)
\left[
\eta^M \partial_M + \frac{1}{2} (-)^M \left(\partial_M\eta^M\right)
\right]
\Psi_{k,\tau}(z)\ .
\end{eqnarray}
Now, let us notice an identity,
\begin{eqnarray}
\label{eq:identity}
\int {\rm d}^6z \Psi_{k,\eta}(z)
\left[
\eta^M \partial_M + \frac{1}{2} (-)^M \left(\partial_M\eta^M\right)
\right]
\Psi_{k,\eta}(z)&=& \frac{1}{2}(-)^M \int {\rm d}^6 z \partial_M 
\left[\Psi_{k,\eta}\left(z\right)\eta^M\Psi_{k,\eta}\left(z\right)\right]\nonumber\\
&=&0\ ,
\end{eqnarray}
where we have used the property that an integration of a total derivative vanishes.
By using this identity several times, the super-trace can be rearranged as
\begin{eqnarray}
{\rm sTr}\,{\cal O}(z',z)  &=& -\frac{1}{2}\int {\rm d}^6z \int \frac{{\rm d}^4 k}{(2\pi)^4} d^2 \tau \left(\Psi_{k,\tau}\left(z\right)+\Psi_{-k,-\tau}\left(z\right)\right)\nonumber\\
&&
\left[
\eta^M \partial_M + \frac{1}{2} (-)^M \left(\partial_M\eta^M\right)
\right]
\left(\Psi_{k,\tau}\left(z\right)+\Psi_{-k,-\tau}\left(z\right)\right)\nonumber\\
&=&
-\frac{1}{4}(-)^M\int {\rm d}^6z \int \frac{{\rm d}^4 k}{(2\pi)^4} d^2 \tau \partial_M
\left[
\left(\Psi_{k,\tau}\left(z\right)+\Psi_{-k,-\tau}\left(z\right)\right)
\eta^M
\left(\Psi_{k,\tau}\left(z\right)+\Psi_{-k,-\tau}\left(z\right)\right)
\right]\nonumber\\
&=&0\ .
\end{eqnarray}
This shows that the measure given in Eq.~(\ref{eq:SD measure}) is actually invariant under the super-Diffeomorphism.
It should be noted that the transformation law in Eq.\,(\ref{eq:QdiffSD}) is crucial to use Eq.\,(\ref{eq:identity}),
and hence, the super-Diffeomorphism invariance does not hold for measures 
with different weights $[D(2{\cal E})^{n}Q]$ $(n \neq 1/2)$.
In fact, the super-Diffeomorphism transformation of  $[D(2{\cal E})^{n}Q]$ $(n \neq 1/2)$
is accompanied by Konishi-Sizuya anomaly~\cite{Konishi:1985tu}.
This argument provides an superfield expression of the arguments in~\cite{Fujikawa:1984qk}.

There is a quicker route to show the super-Differmorphisim invariance of $[DQ_{\rm diff}]$ from 
the very definition of the path-integral measure \cite{Fujikawa:1984qk}.
Let us a consider a superfield $\tilde{Q}(x,\Theta)$ defined in a chiral superspace.
The path-integral measure $[D \tilde{Q}]$ is defined by a Gaussian integration,
\begin{eqnarray}
\label{eq:path integral}
\int [D\tilde{Q}] {\rm exp} \left[ \frac{i}{2} \int {\rm d}^6 z \Theta \tilde{Q} \tilde{Q}  \right] =N\ ,
\end{eqnarray}
where $N$ is a normalization constant.
It should be noted that we have not specified the transformation law of $\tilde Q$ under the super-Diffeomorphism 
at this point.

Next, let us introduce $Q \equiv (2{\cal E})^{-1/2} \tilde{Q}$, and choose
the transformation property of $\tilde{Q}$ as a chiral density multiplet with density weights $1/2$, $Q$ is a chiral scalar multiplet.
Then, from Eq.~(\ref{eq:path integral}), we obtain
\begin{eqnarray}
\int [D \left(2 {\cal E}\right)^{1/2} Q ] {\rm exp} \left[ \frac{i}{2} \int {\rm d}^6 z\, 2{\cal E}\, Q Q  \right] =N\ .
\end{eqnarray}
Now, since $\int {\rm d}^6z {\cal E}QQ$ is invariant under 
the super-Diffeomorphism as $Q$ is the chiral scalar multiplet,
so is the path-integral measure $[D(2{\cal E}^{1/2})Q]$.

In fact, under the super-Diffeomorphism, 
\begin{eqnarray}
{\cal E'} = {\cal E} - \delta_{\rm SD} {\cal E}\ ,~~Q' = Q - \delta_{\rm SD} Q\ ,
\end{eqnarray}
we have the following identities
\begin{eqnarray}
N &=& \int [D \left(2 {\cal E}\right)^{1/2} Q ] {\rm exp} \left[ \frac{i}{2} \int {\rm d}^6 z\, 2{\cal E} \,Q Q  \right]\ , \nonumber\\
&=&\int [D \left(2 {\cal E'}\right)^{1/2} Q' ] {\rm exp} \left[ \frac{i}{2} \int {\rm d}^6 z \,2{\cal E'}\, Q' Q'  \right] \ , \nonumber\\
&=&\int [D \left(2 {\cal E'}\right)^{1/2} Q' ] {\rm exp} \left[ \frac{i}{2} \int {\rm d}^6z  \,2{\cal E}\, Q Q  \right]\ .
\end{eqnarray}
Here, the second equality is just a change of variable.
We have used a super-Diffeomorphism invariance of the exponent in the third equality. 
Thus, from these identities, we find that 
\begin{eqnarray}
[D \left(2 {\cal E'}\right)^{1/2} Q' ] = D [\left(2 {\cal E}\right)^{1/2} Q ]\ ,
\end{eqnarray}
which again shows the super-Diffeomorphism invariance of the measure $[DQ_{\rm diff}]$.
In the same token, we can derive the super-Diffeomorphism invariance of the measure 
of a scalar multiplet $V$ in a real superspace,
\begin{eqnarray}
[D V_{\rm diff}] = [DE^{1/2} V]\ ,
\end{eqnarray}
which we briefly mentioned in Sec.\,\ref{sec:summary}.

\section{Gaugino mass in Pauli-Villars regulalization}
\label{sec:Pauli-Villars}
In this appendix, we show how our method to extract the gaugino mass works in the Pauli-Villar regularization~\cite{Pauli:1949zm}.
In the Pauli-Villar regularization scheme, 
we introduce Pauli-Villars fields, a pair-of fermonic chiral scalar multiplets $P$ and $\bar{P}$ with a unit charge, 
and give them a supersymmetric mass term $\Lambda$ which corresponds to the cutoff scale;
\begin{eqnarray}
{\cal L} = \int {\rm d}^2\Theta\, 2{\cal E}\, \Lambda P \bar{P} + {\rm h.c.}\ .
\end{eqnarray}
As we discussed in Sec.\,\ref{sec:measure}, it is convenient to use the  $F$-type super-Weyl invariant measure, 
$[DQ_{SW}]$, to extract the gaugino mass from the Wilsonian action.
If we also take the measure of the Pauli-Villars fields to be $[DP_{SW}]$, however,
the counter terms appearing when we change the measures are cancelled due to the 
opposite statistic of the  Pauli-Villars fields.
Thus, in this case, the $F$-type super-Weyl invariant measure does not invoke the counter term in Eq.~(\ref{eq:translation}), ${\mit\Delta} S$.

In the absence of ${\mit\Delta} S$, what is the origin of the gaugino mass?
As we discuss in the main text, the gaugino mass is generated only from violations of the approximate $F$-type super-Weyl symmetry.
For a energy scale well below $\Lambda$, the approximate $F$-type super-Weyl symmetry is 
explicitly broken by the mass term of the Pauli-Villars fields.
Thus, the integration of the Pauli-Villars fields generates the gaugino mass, as is discussed in~\cite{Giudice:1998xp}.%
\footnote{More explicitly, the masses of the fermions and the scalars in the Paulli-Villars multiplets
are split by the coupling to $M$ through $\int {\rm d}^2\Theta (2 {\cal E})^{1/3} \Lambda P_{SW} \bar{P}_{SW}$.
}

We can also extract the gaugino mass without explicitly performing the integration of the Pauli-Villars fields.
Well below the mass scale $\Lambda$, a good approximate super-Weyl symmetry is 
the one which is consistent with the mass term of the Pauli-Villars fields.
Thus, the appropriate measures to read off the gaugino mass from the action is 
the combination of $[DQ_{SW}]$ and $[DP_{\rm diff}]$.
With these measures, the counter term is again given by ${\mit\Delta} S$ in Eq.~(\ref{eq:translation}),
from which we can directly read off the anomaly mediated gaugino mass.

\end{document}